\documentclass[10pt,twocolumn,conference]{article}
\usepackage{times}
\usepackage{amsbsy}
\usepackage{latexsym}
\usepackage{amsmath}
\usepackage[ansinew]{inputenc}
\usepackage[dvips ]{graphicx}
\input epsf
\usepackage{epic,eepic}
\usepackage{url}
\date{}

\title{\huge Joint Power Adjustment and Interference Mitigation
Techniques for Cooperative Spread Spectrum Systems \vspace*{0.05em}}

\author{Rodrigo C. de Lamare \\ Communications Research Group \\ Department of Electronics,
    University of York, York Y010 5DD, United Kingdom \\
    Email: \protect\url{rcdl500@ohm.york.ac.uk}
\thanks{\footnotesize The work of Dr R. C. de Lamare was
supported by the University of York, York Y010 5DD, United
Kingdom.   }}

\begin{document}
\maketitle \thispagestyle{empty}

\begin{abstract}
This paper presents joint power allocation and interference
mitigation techniques for the downlink of spread spectrum systems
which employ multiple relays and the amplify and forward
cooperation strategy. We propose a joint constrained optimization
framework that considers the allocation of power levels across the
relays subject to an individual power constraint and the design of
linear receivers for interference suppression. We derive
constrained minimum mean-squared error (MMSE) expressions for the
parameter vectors that determine the optimal power levels across
the relays and the linear receivers. In order to solve the
proposed optimization problem efficiently, we develop joint
adaptive power allocation and interference suppression algorithms
that can be implemented in a distributed fashion. The proposed
stochastic gradient (SG) and recursive least squares (RLS)
algorithms mitigate the interference by adjusting the power levels
across the relays and estimating the parameters of the linear
receiver. SG and RLS channel estimation algorithms are also
derived to determine the coefficients of the channels across the
base station, the relays and the destination terminal. The results
of simulations show that the proposed techniques obtain
significant gains in performance and capacity over non-cooperative
systems and cooperative schemes with equal power allocation.\\

\end{abstract}
%
%

\section{Introduction}

The use of multiple collocated antennas enables the exploitation
of the spatial diversity in wireless channels, mitigating the
effects of fading and enhancing the performance of wireless
communications systems. Unfortunately, due to size and cost it is
often impractical to equip mobile terminals with multiple
antennas. However, spatial diversity gains can be obtained when
terminals with single antennas establish a distributed antenna
array through cooperation \cite{sendonaris}-\cite{laneman04}. In a
cooperative transmission system, terminals or users relay signals
to each other in order to propagate redundant copies of the same
signals to the destination user or terminal. To this end, the
designer must employ a cooperation strategy such as
amplify-and-forward (AF) \cite{laneman04}, decode-and-forward (DF)
\cite{laneman04,huang} and compress-and-forward (CF)
\cite{kramer}.

Recent contributions in the field have considered the problem of
interference mitigation and resource allocation in the context of
cooperative communications with relays \cite{luo}-\cite{chen}.
This problem is of paramount importance in wireless cooperative
cellular, ad-hoc and sensor networks \cite{souryal,fischione} that
utilize spread spectrum systems. Prior work on cooperative
multiuser spread spectrum DS-CDMA systems in interference channels
has not received much attention and has focused on the assessment
of the impact of multiple access interference (MAI) and
intersymbol interference (ISI), the problem of partner selection
\cite{huang,venturino} and the bit error rate (BER) and outage
performance analyzes \cite{vardhe}. Other related contributions on
resource allocation investigated the capacity of ad hoc networks
\cite{comaniciu}, cooperative spatial multiplexing
\cite{levorato}, power and rate control
\cite{kastrinogiannis,chliu}, and scheduling \cite{chen}. There
has been no attempt to jointly consider the problem of resource
allocation and interference mitigation in cooperative multiuser
spread spectrum systems so far.

In this work, we study the downlink of spread spectrum systems
which employ multiple relays and the AF cooperation strategy.
Specifically, we consider the problem of resource allocation and
interference mitigation in multiuser DS-CDMA with a general number
of relays, which have been originally reported in
\cite{delamare_iswcs}. In order to facilitate the receiver design,
we adopt linear multiuser receivers \cite{verdu,delamaretvt} which
only require a training sequence and the timing. More
sophisticated receiver techniques are also possible
\cite{verdu,delamaretc}. We propose a joint constrained
optimization framework that considers the allocation of power
levels among the relays subject to an individual power constraint
and the design of linear receivers. We derive constrained minimum
mean-squared error (MMSE) expressions for the parameter vectors
that determine the optimal power levels across the relays and the
linear receivers. In order to solve the proposed optimization
problem efficiently, we also develop joint adaptive power
allocation and interference suppression algorithms. Specifically,
we derive computationally efficient stochastic gradient (SG) and
recursive least squares (RLS) algorithms that can be employed in a
distributed fashion. The proposed SG and RLS algorithms are
employed to mitigate the effects of the MAI and the ISI, and to
adjust the power levels, increasing the capacity of CDMA networks
with cooperative diversity. These algorithms can be implemented in
a distributed fashion, which means the mobile units compute the
coefficients and employ a low-rate feedback channel to update the
coefficients for the power allocation. In addition, other SG and
RLS algorithms are developed to estimate the parameters of the
channels across the base station, the relays and the destination
terminal of the cooperative DS-CDMA system under consideration.
The proposed algorithms are compared with non-cooperative and
cooperative techniques without power allocation via computer
simulations.


This paper is organized as follows. Section 2 briefly describes a
cooperative DS-CDMA system and data model with multiple relays.
Section 3 is devoted to the problem statement and the constrained
linear MMSE design of the interference mitigation receiver and the
power allocation. Section 4 is dedicated to the derivation of
constrained adaptive SG and RLS algorithms for the estimation of
the parameters of the receiver and the power allocation across the
base station, the relays and the destination terminal. Section 5
is devoted to the development of adaptive channel estimation
algorithms for the cooperative system under consideration. Section
6 presents and discusses the simulation results and Section 7
draws the conclusions of this paper.

\section{System and Data Model}

\begin{figure}[!htb]
\begin{center}
\def\epsfsize#1#2{1.025\columnwidth}
\epsfbox{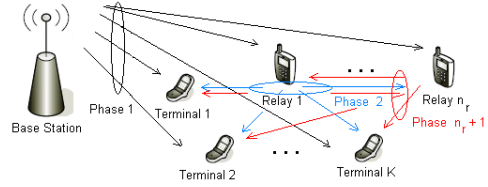} \caption{Downlink of proposed cooperative
multihop DS-CDMA system.} \label{figsys}
\end{center}
\end{figure}

Consider the downlink of a synchronous DS-CDMA system
communicating over multipath channels with QPSK modulation, $K$
users, $N$ chips per symbol and $L$ as the maximum number of
propagation paths for each link. The synchronous DS-CDMA systems
is considered for simplicity as it captures most of the effects of
asynchronous systems with low delay spread
\cite{delamaretvt,delamaretc}. The system is equipped with an AF
protocol that allows communications in multiple hops using $n_r$
relays in a repetitive fashion. It should be remarked that other
cooperation protocols such as DF can be employed without
significant modifications, however, the AF has been adopted for
simplicity and due to its lower complexity for implementation
\cite{laneman04}. We assume that the base station transmits data
organized in packets with $P$ symbols, there is enough training
and control data to coordinate transmissions and cooperation, and
the linear receivers at the terminals are perfectly synchronized.
Since the focus of this work is on the resource allocation and
interference mitigation, we assume perfect synchronization,
however, this assumption can be relaxed in order to account for
more realistic synchronization effects in the network The
cooperative DS-CDMA system under consideration is illustrated in
Fig. \ref{figsys}.

The received signals are filtered by a matched filter, sampled at
the chip rate and organized into $M \times 1$ vectors
${\boldsymbol r}_{bd}[i]$ and ${\boldsymbol r}_{br_i}[i]$, which
describe the signals received from the base station to the
destination and from the base station to the relays, respectively,
as follows
\begin{equation}
\begin{split}
{\boldsymbol r}_{bd}[n] & = \sum_{k=1}^K  a_{bd}^k[n] {\boldsymbol
D}_k {\boldsymbol h}_{bd}[n]b_k[i]  \\ & \quad + {\boldsymbol
\eta}_{bd}[n] +
{\boldsymbol n}_{bd}[n], \\
{\boldsymbol r}_{br_j}[m] & = \sum_{k=1}^K a_{br_j}^k[m]
{\boldsymbol D}_k {\boldsymbol h}_{br_j}[m] {b}_k[i]  \\ & \quad +
{\boldsymbol \eta}_{br_j}[m] +
{\boldsymbol n}_{br_j}[m], \\
& r_j = 1,2, \ldots, n_r,~~ i=0, 1, \ldots, P-1 \\
& n=n_ri+1, ~~ m = n_ri+r_j+1 \label{rvec}
\end{split}
\end{equation}
where $M=N+L-1$, ${\boldsymbol n}_{bd}[i]$ and ${\boldsymbol
n}_{br_j}[i]$ are zero mean complex Gaussian vectors with variance
$\sigma^2$ generated at the receiver of the destination and the
relays, and the vectors ${\boldsymbol \eta}_{bd}[i]$ and
${\boldsymbol \eta}_{br_j}[i]$ represent the intersymbol
interference (ISI).

The $M \times L$ matrix ${\boldsymbol D}_k$ contains versions of
the signature sequences of each user shifted down by one position
at each column as illustrated by
\begin{equation}
{\boldsymbol D}_k = \left[\begin{array}{c c c }
d_{k}(1) &  & {\bf 0} \\
\vdots & \ddots & d_{k}(1)  \\
d_{k}(N) &  & \vdots \\
{\bf 0} & \ddots & d_{k}(N)  \\
 \end{array}\right],
\end{equation}
where ${\boldsymbol d}_k = \big[d_{k}(1), ~d_{k}(2),~ \ldots,~
d_{k}(N) \big]$ stands for the signature sequence of user $k$, the
$L \times 1$ channel vectors from base station to destination,
base station to relay, and relay to destination are ${\boldsymbol
h}_{bd}[n]$, ${\boldsymbol h}_{br_j}[n]$, ${\boldsymbol
h}_{r_jd}[n]$, respectively. By collecting the data vectors in
(\ref{rvec}) (including the links from relays to destination) into
a $(n_r+1)M \times 1$ received vector at the destination we obtain
{
\begin{equation*}
\begin{split}
 \left[\begin{array}{c}
  {\boldsymbol r}_{bd}[n] \\
  {\boldsymbol r}_{r_{1}d}[m] \\
  \vdots \\
  {\boldsymbol r}_{r_{n_r}d}[m]
\end{array}\right] & = \left[\begin{array}{c}
  \sum_{k=1}^K  a_{bd}^k[n] {\boldsymbol D}_k {\boldsymbol h}_{bd}[n]b_k[i] \\
  \sum_{k=1}^K  a_{{r_1}d}^k[m] {\boldsymbol D}_k {\boldsymbol h}_{{r_1}d}[m]{\tilde b}_k^{{r_1}d}[i] \\
  \vdots \\
  \sum_{k=1}^K  a_{{r_{n_r}}d}^k[m] {\boldsymbol D}_k {\boldsymbol h}_{r_{n_r}d}[m]{\tilde b}_k^{{r_{n_r}}d}[i]
\end{array}\right] \\ & \quad + {\boldsymbol \eta}[i] + {\boldsymbol n}[i]
\end{split}
\end{equation*}}
Rewriting the above signals in a compact form yields
\begin{equation}
{\boldsymbol r}[i] = \sum_{k=1}^{K} {\boldsymbol {\mathcal C}}_k
{\boldsymbol {\mathcal H}}[i] {\boldsymbol B}_k[i] {\boldsymbol
a}_k[i] + {\boldsymbol \eta}[i] + {\boldsymbol n}[i],
\end{equation}
where the $(n_r+1)M \times (n_r+1)L$ block diagonal matrix
${\boldsymbol {\mathcal C}}_k$ contains versions of the spreading
sequences of each user, the $(n_r+1)L \times (n_r+1)$ matrix
${\boldsymbol {\mathcal H}}[i]$ contains the channel gains of the
links between the base station and the destination, and the relays
and the destination. The $(n_r+1) \times (n_r+1)$ diagonal matrix
${\boldsymbol B}_k[i] = {\rm diag}(b_k[i]~ {\tilde
b}_k^{{r_1}d}[i] \ldots {\tilde b}_k^{{r_n}d}[i]) $ contains the
symbols transmitted from the base station to the destination
($b_k[i]$) and the $n_r$ symbols transmitted from the relays to
the destination (${\tilde b}_k^{{r_1}d}[i] \ldots {\tilde
b}_k^{{r_n}d}[i]$) on the main diagonal, the $(n_r+1) \times 1$
vector ${\boldsymbol a}_k[i]=[a_{bd}^k[n]~a_{{r_1}d}^k[m]\ldots
a_{{r_{n_r}}d}^k[m]]^T$ of the amplitudes of the links, the
$(n_r+1)M \times 1$ vector ${\boldsymbol \eta}[i]$ with the ISI
terms and $(n_r+1)M \times 1$ vector ${\boldsymbol n}[i]$ with the
noise components at the destination.

\section{Problem Statement and Proposed MMSE Design }

We are interested in jointly designing a linear receiver and
determining the optimal power levels across the relays subject to
an individual power constraint. Let us consider an MMSE approach
for the design of the receiver for user $k$ represented by a
$(n_r+1)M \times 1$ parameter vector ${\boldsymbol w}_k[i]$ and
for the computation of the $(n_r +1) \times 1$ optimal power
allocation vector ${\boldsymbol a}_k[i]$. This problem can be cast
as
\begin{equation}
\begin{split}
[ {\boldsymbol w}_{k,{\rm opt}}[i], {\boldsymbol a}_{k,{\rm
opt}}[i] ] & = \arg \min_{{\boldsymbol w}_k[i], {\boldsymbol
a}_k[i]} ~
E[ |b_k[i] - {\boldsymbol w}_k^H[i]{\boldsymbol r}[i] |^2 ] \\
{\rm subject ~to~} & {\boldsymbol a}_k^H[i] {\boldsymbol a}_k[i] =
P_{A,k} , ~~~  k   = 1,~ 2,~\ldots, ~K. \label{prob}
\end{split}
\end{equation}
The expressions for the parameter vectors ${\boldsymbol w}_k[i]$
and ${\boldsymbol a}_k[i]$ can be obtained by transforming the
above constrained optimization problem into an unconstrained one
with the help of the method of Lagrange multipliers \cite{haykin}
which leads to
\begin{equation}
\begin{split}
{\mathcal L}_{{\rm I}_k} & = E\big[ \big|b_k[i] - {\boldsymbol
w}_k^H[i]\big(\sum_{l=1}^{K} {\boldsymbol {\mathcal C}}_l
{\boldsymbol {\mathcal H}}[i] {\boldsymbol B}_l[i] {\boldsymbol
a}_l[i] + {\boldsymbol \eta}[i] + {\boldsymbol n}[i]\big) \big|^2
\big] \\ & \quad + \lambda ({\boldsymbol a}_k^H[i] {\boldsymbol
a}_k[i] -P_{A,k}), ~~~  k   = 1,~ 2,~\ldots, ~K. \label{lag}
\end{split}
\end{equation}
Fixing ${\boldsymbol a}_k[i]$, computing the gradient terms of the
Lagrangian with respect to ${\boldsymbol w}_{k}[i]$ and equating
them to zero yields
\begin{equation}
{\boldsymbol w}_{k,{\rm opt}}[i] = {\boldsymbol R}^{-1}[i]
{\boldsymbol p}_{{\boldsymbol{\mathcal C}}{\boldsymbol{\mathcal
H}}}[i],~~~  k = 1,~ 2,~\ldots, ~K, \label{wvec}
\end{equation}
where ${\boldsymbol R}[i] = \sum_{k=1}^{K}{\boldsymbol {\mathcal
C}}_k {\boldsymbol {\mathcal H}}[i]{\boldsymbol B}_k[i]
{\boldsymbol a}_k[i] {\boldsymbol a}_k^H[i] {\boldsymbol B}_k^H[i]
{\boldsymbol {\mathcal H}}^H[i] {\boldsymbol {\mathcal C}}_k^H +
\sigma^2 {\boldsymbol I}$ is the covariance matrix and
${\boldsymbol p}_{{\boldsymbol{\mathcal C}}{\boldsymbol{\mathcal
H}}}[i] = E[b_k^*[i] {\boldsymbol r}[i]] = {\boldsymbol {\mathcal
C}}_k {\boldsymbol {\mathcal H}}[i]{\boldsymbol a}_k[i] $ is the
cross-correlation vector. The quantities ${\boldsymbol R}[i]$ and
${\boldsymbol p}_{{\boldsymbol{\mathcal C}}{\boldsymbol{\mathcal
H}}}[i]$ depend on ${\boldsymbol a}_k[i]$. By fixing ${\boldsymbol
w}_k[i]$, computing the gradient terms of the Lagrangian with
respect to ${\boldsymbol a}_{k,{\rm opt}}[i]$ and equating them to
zero, we obtain the following expression for the power allocation
vector
\begin{equation}
{\boldsymbol a}_{k,{\rm opt}}[i] = ( {\boldsymbol R}_{{\boldsymbol
a}_k}[i] + \lambda {\boldsymbol I})^{-1} {\boldsymbol
p}_{{\boldsymbol a}_k}[i], ~~~  k   = 1,~ 2,~\ldots, ~K,
\label{avec}
\end{equation}
where ${\boldsymbol R}_{{\boldsymbol a}_k}[i] = \sum_{k=1}^{K}
{\boldsymbol {\boldsymbol B}}_k^H[i] {\boldsymbol {\mathcal
H}}^H[i] {\boldsymbol {\mathcal C}}_k^H {\boldsymbol w}_{k}[i]
{\boldsymbol w}_{k}^H[i]{\boldsymbol {\mathcal C}}_k {\boldsymbol
{\mathcal H}}[i]{\boldsymbol B}_k[i]$ is the $(n_r+1) \times
(n_r+1)$ covariance matrix and the $(n_r+1) \times 1$
cross-correlation vector is ${\boldsymbol p}_{{\boldsymbol
a}_k}[i] = E[b_k[i] {\boldsymbol B}_k^H[i] {\boldsymbol {\mathcal
H}}[i]^H {\boldsymbol {\mathcal C}}_k^H {\boldsymbol w}_k[i] ]$.

The expressions in (\ref{wvec}) and (\ref{avec}) depend on each
other and require the estimation of the channel matrix
${\boldsymbol {\mathcal H}}[i]$, which is identical for each user
as we are dealing with a downlink channel. The expressions in
(\ref{wvec}) and (\ref{avec}) require matrix inversions with cubic
complexity ( $O(((n_r+1)M)^3)$ and $O((n_r+1)^3)$, should be
iterated as they depend on each other and  on user $k$. It should
be remarked that the proposed optimization problem is non-convex
and may present multiple solutions due to the joint optimization
of parameters. However, the experience with the proposed
algorithms suggest that the solutions may be identical because we
did not notice problems with local minima or loss of performance
under different initialization. A study of the optimization
problem is beyond the scope of this work but seems to be an
interesting topic for future investigation. In what follows, we
will develop algorithms for computing ${\boldsymbol a}_{k,{\rm
opt}}[i]$, ${\boldsymbol w}_{k,{\rm opt}}[i]$ and estimating the
channel matrix ${\boldsymbol {\mathcal H}}[i]$.

\section{Proposed Constrained Estimation Algorithms for
Receiver Design and Power Allocation}

In this section we present adaptive constrained SG and RLS
estimation algorithms to estimate the parameters of the linear
receiver and the power allocation. A key feature of the proposed
algorithms is that they can be employed in a distributed fashion.
The only information that needs to be sent via a feedback channel
is the power allocation vector.

\subsection{Adaptive Constrained Estimation and Power Allocation with SG Algorithms}

In this subsection, we will develop SG algorithms for computing
$\hat{\boldsymbol w}_k[i]$ and $\hat{\boldsymbol a}[i]$
recursively. Let us consider the proposed constrained optimization
in (\ref{prob}), resort to the method of Lagrange multipliers
\cite{haykin} and express the following Lagrangian
\begin{equation}
\begin{split}
{\mathcal L} & = E\bigg[ |b_k[i] - \hat{\boldsymbol
w}_k^H[i]\bigg(\sum_{l=1}^{K} {\boldsymbol {\mathcal C}}_l
\hat{\boldsymbol {\mathcal H}}[i] {\boldsymbol B}_l[i]
\hat{\boldsymbol a}_l[i] + {\boldsymbol \eta}[i] + {\boldsymbol
n}[i]\bigg) |^2 \bigg] \\ & \quad + \lambda (\hat{\boldsymbol
a}_k^H[i] \hat{\boldsymbol a}_k[i] -P_{A,k}), \label{lag2}
\end{split}
\end{equation}
where $\hat{\boldsymbol w}_k[i]$, $\hat{\boldsymbol {\mathcal
H}}[i]$, and $\hat{\boldsymbol a}[i]$ are parameter estimates of
the receiver, the channel and the power allocation to be
determined. Due to the nature of the problem, we need to jointly
estimate these parameters. To this end, we will develop joint SG
algorithms that can perform this task with low complexity.

We consider the Lagrangian in (\ref{lag2}) and calculate the
instantaneous gradient terms of it with respect to
$\hat{\boldsymbol w}_k[i]$, and $\hat{\boldsymbol a}[i]$,
respectively, as follows:
\begin{equation}
\begin{split}
\nabla{\mathcal L}_{\hat{\boldsymbol w}_k^*[i]} & = -
\bigg(\sum_{l=1}^{K} {\boldsymbol {\mathcal C}}_l \hat{\boldsymbol
{\mathcal H}}[i] {\boldsymbol B}_l[i] \hat{\boldsymbol a}_l[i] +
{\boldsymbol \eta}[i] + {\boldsymbol n}[i]\bigg) \cdot
\bigg(b_k[i] \\ & \quad - \hat{\boldsymbol
w}_k^H[i]\bigg(\sum_{l=1}^{K} {\boldsymbol {\mathcal C}}_l
\hat{\boldsymbol {\mathcal H}}[i] {\boldsymbol B}_l[i]
\hat{\boldsymbol a}_l[i] + {\boldsymbol \eta}[i] + {\boldsymbol
n}[i]\bigg)^* \\ & = - {\boldsymbol r}[i] e^*[i],
\end{split}
\end{equation}
\begin{equation}
\begin{split}
\nabla{\mathcal L}_{\hat{\boldsymbol a}_k^*[i]} & = - {\boldsymbol
B}_k^H[i] \hat{\boldsymbol{\mathcal H}}^H[i] {\boldsymbol
{\mathcal C}}_k^H  \hat{\boldsymbol w}_k[i]  \bigg(b_k[i] \\ &
\quad - \hat{\boldsymbol w}_k^H[i]\bigg(\sum_{l=1}^{K}
{\boldsymbol {\mathcal C}}_l \hat{\boldsymbol {\mathcal H}}[i]
{\boldsymbol B}_l[i] \hat{\boldsymbol a}_l[i] + {\boldsymbol
\eta}[i] + {\boldsymbol n}[i]\bigg) + \lambda \hat{\boldsymbol
a}_k[i] \\ & = - {\boldsymbol B}_k^H[i] \hat{\boldsymbol{\mathcal
H}}^H[i] {\boldsymbol {\mathcal C}}_k^H \hat{\boldsymbol w}_k[i]
e[i] + \lambda \hat{\boldsymbol a}_k[i],
\end{split}
\end{equation}
where $e[i] = b_k[i] - \hat{\boldsymbol w}_k^H[i]{\boldsymbol
r}[i]  = b_k[i] - \hat{\boldsymbol w}_k^H[i]\big(\sum_{l=1}^{K}
{\boldsymbol {\mathcal C}}_l \hat{\boldsymbol {\mathcal H}}[i]
{\boldsymbol B}_l[i] \hat{\boldsymbol a}_l[i] + {\boldsymbol
\eta}[i] + {\boldsymbol n}[i]\big)$ is the error signal, which is
a function of $\hat{\boldsymbol w}_k[i]$, $\hat{\boldsymbol
{\mathcal H}}[i]$, and $\hat{\boldsymbol a}[i]$.

Adaptive SG algorithms can be developed by using the expressions
for the instantaneous gradients above and using them with SG
descent rules \cite{haykin}, yielding
\begin{equation}
\begin{split}
\hat{\boldsymbol w}_k[i+1] & = \hat{\boldsymbol w}_k[i] - \mu
\nabla{\mathcal L}_{\hat{\boldsymbol w}_k^*[i]} \\ & =
\hat{\boldsymbol w}_k[i] + \mu e^*[i] {\boldsymbol r}[i],
\label{sgrecw}
\end{split}
\end{equation}
\begin{equation}
\begin{split}
\hat{\boldsymbol a}_k[i+1] & = \hat{\boldsymbol a}_k[i] - \alpha
\nabla{\mathcal L}_{\hat{\boldsymbol a}_k^*[i]} \\ & =
\hat{\boldsymbol a}_k[i] + \alpha \big(  {\boldsymbol B}_k^H[i]
\hat{\boldsymbol{\mathcal H}}^H[i] {\boldsymbol {\mathcal C}}_k^H
\hat{\boldsymbol w}_k[i] e[i] + \lambda \hat{\boldsymbol a}_k[i]
\big), \label{sgreca}
\end{split}
\end{equation}
where $\mu$ and $\alpha$ are the step sizes for the recursions for
the receiver and the power allocation, respectively. Notice that
in the recursion for computing $\hat{\boldsymbol a}_k[i]$, the
designer needs to determine the value of the Lagrange multiplier
$\lambda$. There are two alternative approaches to that. The first
is to substitute (\ref{sgreca}) into the constraint
$\hat{\boldsymbol a}_k^H[i+1] \hat{\boldsymbol a}_k[i+1] =
P_{A,k}$ and then solve the following quadratic equation:
\begin{equation}
a \lambda^2 + b \lambda + c = 0,
\end{equation}
where the coefficients of the equation are
\begin{equation}
\begin{split}
a & = P_{A,k}, \\
b & = 2 P_{A,k} - \alpha^2 ( \hat{\boldsymbol w}_k^H[i]
{\boldsymbol {\mathcal C}}_k \hat{\boldsymbol {\mathcal H}}[i]
{\boldsymbol B}_k[i] \hat{\boldsymbol a}_k[i] \\ & \quad +
\hat{\boldsymbol a}_k^H[i]{\boldsymbol B}_k^H[i]
\hat{\boldsymbol{\mathcal H}}^H[i] {\boldsymbol {\mathcal C}}_k^H
\hat{\boldsymbol w}_k[i]),\\
c & = \alpha( \hat{\boldsymbol a}_k^H[i]{\boldsymbol B}_k^H[i]
\hat{\boldsymbol{\mathcal H}}^H[i] {\boldsymbol {\mathcal C}}_k^H
\hat{\boldsymbol w}_k[i] e[i]  + \hat{\boldsymbol w}_k^H[i]
{\boldsymbol {\mathcal C}}_k \hat{\boldsymbol {\mathcal H}}[i]
{\boldsymbol B}_k[i] \hat{\boldsymbol a}_k[i]) \\ & \quad +
\alpha^2 ( \hat{\boldsymbol w}_k^H[i] {\boldsymbol {\mathcal C}}_k
\hat{\boldsymbol {\mathcal H}}[i] {\boldsymbol B}_k[i]{\boldsymbol
B}_k^H[i] \hat{\boldsymbol{\mathcal H}}^H[i] {\boldsymbol
{\mathcal C}}_k^H \hat{\boldsymbol w}_k[i] |e[i]|^2 ). \nonumber
\end{split}
\end{equation}
The solutions of this quadratic equation are
\begin{equation}
\begin{split}
\lambda_1 & = (-b-b\sqrt{b^2 - 4 ac})/2a, \\
\lambda_2 & = (-b+b\sqrt{b^2 - 4 ac})/2a. \nonumber \end{split}
\end{equation}
These solutions have to be computed for every time instant $i$ and
checked before substituting them into (\ref{sgreca}).

The second approach to computing the power allocation and ensuring
the constraint is as follows. The constraint is relaxed at first
by making $\lambda = 0$ and performing the following recursion:
\begin{equation}
\begin{split}
\hat{\boldsymbol a}_k[i+1] & = \hat{\boldsymbol a}_k[i] + \alpha
\big(  {\boldsymbol B}_k^H[i] \hat{\boldsymbol{\mathcal H}}^H[i]
{\boldsymbol {\mathcal C}}_k^H \hat{\boldsymbol w}_k[i] e[i]
\big). \label{sgreca2}
\end{split}
\end{equation}
This is followed by a procedure to enforce the individual power
constraint for each user $k$, i.e., $\hat{\boldsymbol
a}_k^H[i+1]\hat{\boldsymbol a}_k[i+1] = P_{A,k}$, which is
described by
\begin{equation}
\hat{\boldsymbol a}_k[i+1] \leftarrow P_{A,k} ~ \hat{\boldsymbol
a}_k[i+1] \Big(\sqrt{\hat{\boldsymbol a}_k^H[i+1]\hat{\boldsymbol
a}_k[i+1]}\Big)^{-1} \label{sgreca2x}
\end{equation}
The algorithms for recursive computation of $\hat{\boldsymbol
w}_k[i]$ and $\hat{\boldsymbol a}_k[i]$ require estimates of the
channel matrix ${\boldsymbol {\mathcal H}}[i]$, which will also be
developed in what follows. A comparison between the two approaches
for deriving SG algorithms will be illustrated via simulations.
The complexity of the proposed algorithm is $O((n_r+1)M)$ for
calculating $\hat{\boldsymbol w}_k[i]$ and $O((n_r+1)^2M)$ for
obtaining $\hat{\boldsymbol a}_k[i]$.

\subsection{Adaptive Constrained Estimation and Power Allocation with RLS Algorithms}

Specifically, we consider the problem of the previous section
using an exponentially weighted least squares criterion and
develop an RLS algorithm for the proposed task. Let us now
consider the following proposed least squares (LS) optimization
problem
\begin{equation}
\begin{split}
[ \hat{\boldsymbol w}_{k}[i], \hat{\boldsymbol a}_{k} [i] ] & =
\arg \min_{{\boldsymbol w}_k[i], {\boldsymbol a}_k[i]} ~
\sum_{l=1}^{i} \alpha^{i-l} |b_k[l] - {\boldsymbol w}_k^H[i]{\boldsymbol r}[l] |^2  \\
{\rm subject ~to~} & {\boldsymbol a}_k^H[i] {\boldsymbol a}_k[i] =
P_{A,k}, ~~~ {\rm for}~~ k   = 1,~ 2,~\ldots, ~K, \label{probls}
\end{split}
\end{equation}
where $\alpha$ is a forgetting factor. The LS expressions for the
parameter vectors $\hat{\boldsymbol w}_k[i]$ and $\hat{\boldsymbol
a}_k[i]$ can be obtained in a similar way to the previous section
and are given for each user by
\begin{equation}
\hat{\boldsymbol w}_{k}[i] = \hat{\boldsymbol R}^{-1}[i]
\hat{\boldsymbol p}_{{\boldsymbol{\mathcal
C}}{\boldsymbol{\mathcal H}}}[i] \label{wvecls}
\end{equation}
\begin{equation}
\hat{\boldsymbol a}_{k}[i] = ( \hat{\boldsymbol R}_{{\boldsymbol
a}_k}[i] + \lambda {\boldsymbol I})^{-1} \hat{\boldsymbol
p}_{{\boldsymbol a}_k}[i] \label{avecls}
\end{equation}
where $\hat{\boldsymbol R}[i]  = \sum_{l=1}^i \alpha^{l=i}
{\boldsymbol r}[l]{\boldsymbol r}^H[l] $ is the estimate of the
covariance matrix and $\hat{\boldsymbol p}_{{\boldsymbol{\mathcal
C}}{\boldsymbol{\mathcal H}}}[i] = \sum_{l=1}^i \alpha^{l=i}
b_k^*[l] {\boldsymbol r}[l] $ is the estimate of the
cross-correlation vector,  $\hat{\boldsymbol R}_{{\boldsymbol
a}_k}[i] = \sum_{l=1}^i \alpha^{l=i} {\boldsymbol {\boldsymbol
B}}_k^H[l] {\boldsymbol {\mathcal H}}^H[l] {\boldsymbol {\mathcal
C}}_k^H \hat{\boldsymbol w}_k[l] \hat{\boldsymbol
w}_k^H[l]{\boldsymbol {\mathcal C}}_k {\boldsymbol {\mathcal
H}}[l]{\boldsymbol B}_k[l] $ and $\hat{\boldsymbol
p}_{{\boldsymbol a}_k}[i] = \sum_{l=1}^i \alpha^{l=i} b_k[l]
{\boldsymbol B}_k^H[l] {\boldsymbol {\mathcal H}}[l]^H
{\boldsymbol {\mathcal C}}_k^H \hat{\boldsymbol w}_k[l] $. The
quantity $\lambda$ is the Lagrange multiplier and also plays the
role of regularization term. Due to the difficulty of obtaining a
closed form for this parameter, we will rely on a numerical
solution for obtaining an appropriate value for it. The
expressions in (\ref{wvecls}) and (\ref{avecls}) require matrix
inversions with cubic complexity ( $O(((n_r+1)M)^3)$ and
$O((n_r+1)^3)$, should be iterated as they depend on each other
and still require channel estimates.

Our goal now is to obtain a recursive solution to the expressions
in (\ref{wvecls}) and (\ref{avecls}) and reduce the required
computations. To this end, we will resort to the theory of
adaptive algorithms \cite{haykin} and derive a constrained joint
iterative recursive least squares (RLS) algorithm. This algorithm
will compute $\hat{\boldsymbol w}_k[i]$ and $\hat{\boldsymbol
a}_k[i]$ and will exchange information between the recursion for
improved performance. In order to develop the algorithm, we fix
$\hat{\boldsymbol a}_k[i]$ and compute the inverse of
$\hat{\boldsymbol R}[i]$ using the matrix inversion lemma
\cite{haykin} to obtain $\hat{\boldsymbol w}_k[i]$. If we define
${\boldsymbol \Phi}[i] = \hat{\boldsymbol R}[i]$ then we can
obtain the recursions
\begin{equation}
{\boldsymbol k}[i] = \frac{\alpha^{-1} {\boldsymbol \Phi}[i]
{\boldsymbol r}[i]}{1+ \alpha^{-1} {\boldsymbol r}^H[i]
{\boldsymbol \Phi}[i] {\boldsymbol r}[i]} \label{mil1a}
\end{equation}
and
\begin{equation}
{\boldsymbol \Phi}[i] = \alpha^{-1} {\boldsymbol \Phi}[i-1] -
\alpha^{-1} {\boldsymbol k}[i] {\boldsymbol r}^H[i] {\boldsymbol
\Phi}[i-1] \label{mil1b}
\end{equation}
Using the LS expression in (\ref{wvecls}) and the recursion
$\hat{\boldsymbol p}_{{\boldsymbol{\mathcal
C}}{\boldsymbol{\mathcal H}}}[i] = \alpha {\boldsymbol
p}_{{\boldsymbol{\mathcal C}}{\boldsymbol{\mathcal H}}}[i-1] +
b_k^*[i] {\boldsymbol r}[i]$ we get
\begin{equation}
\hat{\boldsymbol w}_{k}[i] = \hat{\boldsymbol R}^{-1}[i]
\hat{\boldsymbol p}_{{\boldsymbol{\mathcal
C}}{\boldsymbol{\mathcal H}}}[i]  = \alpha {\boldsymbol
\Phi}[i]{\boldsymbol p}_{{\boldsymbol{\mathcal
C}}{\boldsymbol{\mathcal H}}}[i-1] +  {\boldsymbol \Phi}[i]
{\boldsymbol r}[i] b_k^*[i] \label{wvecrls1}
\end{equation}
Using the expression in (\ref{mil1b}) for ${\boldsymbol \Phi}[i]$,
substituting in (\ref{wvecrls1}) and manipulating the terms yields
\begin{equation}
{\boldsymbol w}_k[i] = {\boldsymbol w}_k[i-1] + {\boldsymbol k}[i]
\xi^*[i] \label{wrls}
\end{equation}
where the \textit{a priori} estimation error is given by
\begin{equation}
\xi[i] = b_k[i] - {\boldsymbol w}_k^H[i-1] {\boldsymbol r}[i].
\label{ape1}
\end{equation}
The derivation for the recursion that estimates the power
allocation follows a similar approach to the computation of
$\hat{\boldsymbol w}_k[i]$. However, there are some difficulties
related to the enforcement of the constraint and how to
incorporate it into an efficient RLS algorithm. At this point, a
modification is required in order to complete the derivation of
the proposed RLS algorithm. This is because the LS expression in
(\ref{avecls}) incorporates a Lagrange multiplier ($\lambda$) to
ensure the individual power constraint, which is difficult to
embed within the matrix inversion lemma. Our approach is to obtain
the LS expression for the problem in (\ref{probls}) without the
constraint and then ensure the constraint is incorporated via a
subsequent normalization procedure. In order to develop the
recursions for $\hat{\boldsymbol a}_k[i]$, we need to compute the
inverse of $\hat{\boldsymbol R}_{\boldsymbol{a}_k}[i]$. To this
end, let us first define ${\boldsymbol \Phi}_{\boldsymbol {a}_k} =
\hat{\boldsymbol R}_{\boldsymbol{a}_k}[i]$ and employ the matrix
inversion lemma \cite{haykin} as follows:
\begin{equation}
{\boldsymbol k}_{\boldsymbol{a}_k}[i] = \frac{\alpha^{-1}
{\boldsymbol \Phi}_{\boldsymbol {a}_k}[i] {\boldsymbol
B}_k^H[i]\hat{\boldsymbol{\mathcal H}}^H[i]{\boldsymbol {\mathcal
C}}_k^H \hat{\boldsymbol w}_k[i]}{1+ \alpha^{-1} \hat{\boldsymbol
w}_k^H[i]{\boldsymbol {\mathcal C}}_k \hat{\boldsymbol{\mathcal
H}}[i]{\boldsymbol B}_k[i] {\boldsymbol \Phi}_{\boldsymbol
{a}_k}[i] {\boldsymbol B}_k^H[i]\hat{\boldsymbol{\mathcal
H}}^H[i]{\boldsymbol {\mathcal C}}_k^H \hat{\boldsymbol w}_k[i]}
\label{mil2a}
\end{equation}
and
\begin{equation}
{\boldsymbol \Phi}_{\boldsymbol {a}_k}[i] = \alpha^{-1}
{\boldsymbol \Phi}_{\boldsymbol {a}_k}[i-1] - \alpha^{-1}
{\boldsymbol k}_{\boldsymbol{a}_k}[i] \hat{\boldsymbol
w}_k^H[i]{\boldsymbol {\mathcal C}}_k \hat{\boldsymbol{\mathcal
H}}[i]{\boldsymbol B}_k[i] {\boldsymbol \Phi}_{\boldsymbol
{a}_k}[i-1] \label{mil2b}
\end{equation}
Now a recursive equation for computing $\hat{\boldsymbol
p}_{{\boldsymbol a}_k}[i]$ can be devised by relying on time
averages as given by
\begin{equation}
\hat{\boldsymbol p}_{{\boldsymbol a}_k}[i] = \alpha
\hat{\boldsymbol p}_{{\boldsymbol a}_k}[i-1] + b_k[i]{\boldsymbol
B}_k^H[i]\hat{\boldsymbol{\mathcal H}}^H[i]{\boldsymbol {\mathcal
C}}_k^H \hat{\boldsymbol w}_k[i] \label{pavec}
\end{equation}
 The
substitution of (\ref{pavec}) into the unconstrained LS expression
yields
\begin{equation}
\begin{split}
\hat{\boldsymbol a}_{k}[i] & = \hat{\boldsymbol
R}_{\boldsymbol{a}_k}^{-1}[i] \hat{\boldsymbol p}_{{\boldsymbol
a}_k}[i] \\ & = \alpha {\boldsymbol \Phi}_{\boldsymbol {a}_k}[i]
\hat{\boldsymbol p}_{{\boldsymbol a}_k}[i-1] + b_k[i] {\boldsymbol
\Phi}_{\boldsymbol {a}_k}[i]{\boldsymbol
B}_k^H[i]\hat{\boldsymbol{\mathcal H}}^H[i]{\boldsymbol {\mathcal
C}}_k^H \hat{\boldsymbol w}_k[i] \label{aveuls}
\end{split}
\end{equation}
Substituting (\ref{mil2b}) into the above expression and after
some algebraic manipulations with the terms we obtain
\begin{equation}
\begin{split}
\hat{\boldsymbol a}_{k}[i] &  =  {\boldsymbol \Phi}_{\boldsymbol
{a}_k}[i-1] \hat{\boldsymbol p}_{{\boldsymbol a}_k}[i-1]  \\ &
\quad- {\boldsymbol k}_{\boldsymbol {a}_k}[i] {\boldsymbol
w}_k^H[i]{\boldsymbol {\mathcal C}}_k \hat{\boldsymbol{\mathcal
H}}[i]{\boldsymbol B}_k[i] {\boldsymbol \Phi}_{\boldsymbol
{a}_k}[i] \hat{\boldsymbol p}_{{\boldsymbol a}_k}[i-1] +
{\boldsymbol k}_{\boldsymbol {a}_k}b_k^*[i] \label{aveuls}
\end{split}
\end{equation}
By further manipulating the above expressions we get the recursive
equation for $\hat{\boldsymbol a}_k[i]$
\begin{equation}
\hat{\boldsymbol a}_k[i] = \hat{\boldsymbol a}_k[i] + {\boldsymbol
k}_{\boldsymbol{a}_k}[i] \xi_{\boldsymbol{a}_k}^*[i]
\end{equation}
where the \textit{a priori} estimation error for this recursion is
\begin{equation}
\xi_{\boldsymbol{a}_k}[i] = b_k[i] - \hat{\boldsymbol
a}_k^H[i]{\boldsymbol B}_k^H[i]\hat{\boldsymbol{\mathcal
H}}^H[i]{\boldsymbol {\mathcal C}}_k^H \hat{\boldsymbol w}_k[i]
\end{equation}
In order to ensure the individual power constraint ${\boldsymbol
a}_k^H[i]{\boldsymbol a}_k[i] = P_{A,k}$, we apply the following
rule
\begin{equation}
\hat{\boldsymbol a}_k[i] \leftarrow P_{A,k} ~ \hat{\boldsymbol
a}_k[i] \Big(\sqrt{\hat{\boldsymbol a}_k^H[i]\hat{\boldsymbol
a}_k[i]}\Big)^{-1}
\end{equation}
The algorithms for recursive computation of $\hat{\boldsymbol
w}_k[i]$ and ${\boldsymbol a}_k[i]$ require estimates of the
channel vector ${\boldsymbol {\mathcal H}}[i]$, which will also be
developed in what follows. The complexity of the proposed
algorithm is $O(((n_r+1)M)^2)$ for calculating $\hat{\boldsymbol
w}_k[i]$ and $O((n_r+1)^2)$ for obtaining $\hat{\boldsymbol
a}_k[i]$.

\section{Adaptive Channel Estimation Algorithms}

In this section, we describe adaptive SG and RLS channel
estimation algorithms for the cooperative DS-CDMA system operating
with the AF cooperation protocol considered in Section 2. The
proposed algorithms are developed for use with the parameter
estimation algorithms derived in the previous section for receiver
design and power allocation.

\subsection{Adaptive SG Channel Estimation}

In this part we present an adaptive SG channel estimation
algorithm for determining the parameters of the channels across
the links comprising the base station, the relays and the
destination terminal. In order to derive such channel estimator,
we first cast it as the following optimization problem
\begin{equation}
\begin{split}
\hat{\boldsymbol {\mathcal H}}[i] & = \arg \min_{{\boldsymbol
{\mathcal H}}[i]} ~ E[ || {\boldsymbol r}[i] - {
b}_k[i]{\boldsymbol {\mathcal C}}_k \hat{\boldsymbol {\mathcal
H}}[i] \hat{\boldsymbol a}_k[i] ||^2]. \label{cestprob2}
\end{split}
\end{equation}
In order to derive an SG channel estimation algorithm, we start
with the description of a cost function associated with the
optimization problem in (\ref{cestprob2}) described by:
\begin{equation}
\begin{split}
{\mathcal C} & = E[ || {\boldsymbol r}[i] - { b}_k[i]{\boldsymbol
{\mathcal C}}_k \hat{\boldsymbol {\mathcal H}}[i] \hat{\boldsymbol
a}_k[i] ||^2] \\
& = E[ ({\boldsymbol r}[i] - { b}_k[i]{\boldsymbol {\mathcal C}}_k
\hat{\boldsymbol {\mathcal H}}[i] \hat{\boldsymbol
a}_k[i])^H({\boldsymbol r}[i] - { b}_k[i]{\boldsymbol {\mathcal
C}}_k \hat{\boldsymbol {\mathcal H}}[i] \hat{\boldsymbol
a}_k[i])]\\ & = E[ {\boldsymbol r}^H[i]{\boldsymbol r}[i]  - {
b}_k^*[i]\hat{\boldsymbol a}_k^H[i] \hat{\boldsymbol {\mathcal
H}}^H[i] {\boldsymbol {\mathcal C}}_k^H{\boldsymbol r}[i] \\ &
\quad - { b}_k[i] {\boldsymbol r}^H[i] {\boldsymbol {\mathcal
C}}_k \hat{\boldsymbol {\mathcal H}}[i] \hat{\boldsymbol a}_k[i] -
\hat{\boldsymbol a}_k^H[i] \hat{\boldsymbol {\mathcal H}}^H[i]
{\boldsymbol {\mathcal C}}_k^H{\boldsymbol {\mathcal C}}_k
\hat{\boldsymbol {\mathcal H}}[i] \hat{\boldsymbol a}_k[i]]  .
\label{costcest2}
\end{split}
\end{equation}
Computing the gradient terms of (\ref{costcest2}) with respect to
the $(n_r+1)L \times (n_r+1)$ channel estimate matrix
$\hat{\boldsymbol {\mathcal H}}[i]$, we get
\begin{equation}
\begin{split}
\nabla {\mathcal C}_{\hat{\boldsymbol {\mathcal H}}^*[i]} & = -
{\boldsymbol {\mathcal C}}_k^H {\boldsymbol r}[i] \hat{\boldsymbol
a}_k^H[i] b_k^*[i] + {\boldsymbol {\mathcal C}}_k^H {\boldsymbol
{\mathcal C}}_k \hat{\boldsymbol {\mathcal H}}[i] \hat{\boldsymbol
a}_k[i] \hat{\boldsymbol a}_k^H[i]
\end{split}
\end{equation}
Using the above result on the gradient of the cost function and
resorting to a SG optimization recursion, we obtain
\begin{equation}
\begin{split}
\hat{\boldsymbol {\mathcal H}}[i+1] & = \hat{\boldsymbol {\mathcal
H}}[i] - \nu \nabla {\mathcal C}_{\hat{\boldsymbol {\mathcal
H}}[i]} \\ & = \hat{\boldsymbol {\mathcal H}}[i] + \nu
({\boldsymbol {\mathcal C}}_k^H {\boldsymbol {\mathcal C}}_k
\hat{\boldsymbol {\mathcal H}}[i] \hat{\boldsymbol a}_k[i]
\hat{\boldsymbol a}_k^H[i] \\ & \quad - {\boldsymbol {\mathcal
C}}_k^H {\boldsymbol r}[i] \hat{\boldsymbol a}_k^H[i] b_k^*[i] ),
\end{split}
\end{equation}
where $\nu$ is a step size. This SG algorithm for channel
estimation works very well and can accurately determine the
coefficients of the channels across the links comprising the base
station, the relays and the destination terminal. The complexity
of the proposed SG channel estimation algorithm is
$O(((n_r+1)ML))$.

\subsection{Adaptive RLS Channel Estimation}

In this part we present an adaptive RLS channel estimation
algorithm for determining the parameters of the channels across
the links comprising the base station, the relays and the
destination terminal. In order to derive such channel estimator,
we first cast it as the following optimization problem
\begin{equation}
\begin{split}
\hat{\boldsymbol {\mathcal H}}[i] & = \arg \min_{{\boldsymbol
{\mathcal H}}[i]} ~ \sum_{l=1}^{i} \alpha^{i-l} || {\boldsymbol
r}[l] - {\boldsymbol {\mathcal C}}_k \hat{\boldsymbol {\mathcal
H}}[i] { b}_k[l] \hat{\boldsymbol a}_k[l] ||^2
\end{split}
\end{equation}
Due to the structure of the $(n_r+1)L \times (n_r+1)$ channel
matrix $\hat{\boldsymbol {\mathcal H}}[i]$ and its relationship
with the power allocation vector $\hat{\boldsymbol a}_k[l]$, we
found that it is convenient for the derivation and algorithm
development to combine them into a $(n_r+1)L \times 1$ channel
estimate vector
\begin{equation}
\hat{\tilde{\boldsymbol h}}[i] = \hat{\boldsymbol {\mathcal H}}[i]
 \hat{\boldsymbol a}_k[l] \label{rel}
\end{equation}
and recast the optimization problem as
\begin{equation}
\begin{split}
\hat{\tilde{\boldsymbol h}}[i] & = \arg \min_{{\tilde{\boldsymbol
h}}[i]} ~ \sum_{l=1}^{i} \alpha^{i-l} || {\boldsymbol r}[l] -
b_k[l] {\boldsymbol {\mathcal C}}_k \hat{\tilde{\boldsymbol h}}[i]
||^2
\end{split}
\end{equation}
The solution to the above optimization problem is given by
\begin{equation}
\hat{\tilde{\boldsymbol h}}[i] = {\boldsymbol \Phi}_{\boldsymbol
h} {\boldsymbol p}_{\boldsymbol {h}}[i]  \label{cest}
\end{equation}
where ${\boldsymbol \Phi}_{\boldsymbol h} = ({\boldsymbol
{\mathcal C}}_k^H{\boldsymbol {\mathcal C}}_k)^{-1}$ and
${\boldsymbol p}_{\boldsymbol {h}}[i] = \sum_{l=1}^{i}
\alpha^{i-l} b_k^*[l] {\boldsymbol {\mathcal C}}_k^H {\boldsymbol
r}[l]$. It should be remarked that the matrix inversion in
${\boldsymbol \Phi}_{\boldsymbol h}$ can pre-computed and stored
at the receiver for systems with repetitive spreading codes. In
order to develop a recursive algorithm for estimating the channel,
we express ${\boldsymbol p}_{\boldsymbol {h}}[i]$ via the
following recursion
\begin{equation}
{\boldsymbol p}_{\boldsymbol {h}}[i] = \alpha{\boldsymbol
p}_{\boldsymbol {h}}[i-1] + b_k^*[i]  {\boldsymbol {\mathcal
C}}_k^H {\boldsymbol r}[i]
\end{equation}
Substituting it into (\ref{cest}) we obtain
\begin{equation}
\begin{split}
\hat{\tilde{\boldsymbol h}}[i] & = \alpha \hat{\tilde{\boldsymbol
h}}[i-1] + b_k^*[i]  {\boldsymbol \Phi}_{\boldsymbol h}
{\boldsymbol {\mathcal C}}_k^H {\boldsymbol r}[i] \label{crlsest}
\end{split}
\end{equation}
Once $\hat{\tilde{\boldsymbol h}}[i]$ is computed, we need to
apply a transformation in order to obtain $\hat{\boldsymbol
{\mathcal H}}[i]$. This is carried out by manipulating
algebraically the relation in (\ref{rel}) with the post
multiplication of $\hat{\boldsymbol a}_k^H[l]$, which yields
\begin{equation}
\begin{split}
\hat{\tilde{\boldsymbol h}}[i]\hat{\boldsymbol a}_k^H[l] & =
\hat{\boldsymbol {\mathcal H}}[i]  \hat{\boldsymbol a}_k[l]
\hat{\boldsymbol a}_k^H[l] \label{alg}
\end{split}
\end{equation}
Now computing the Moore-Penrose pseudo-inverse of
$\hat{\boldsymbol a}_k[l] \hat{\boldsymbol a}_k^H[l] $ we obtain
the following relation
\begin{equation}
\hat{\boldsymbol {\mathcal H}}[i] = \hat{\tilde{\boldsymbol h}}[i]
\hat{\boldsymbol a}_k^H[l] (\hat{\boldsymbol a}_k[l]
\hat{\boldsymbol a}_k^H[l])^{\dag}
\end{equation}
where $(\cdot)^{\dag}$ denotes Moore-Penrose pseudo-inverse
\cite{haykin}. This procedure for channel estimation also works
well and can accurately determine the coefficients of the channels
across the links comprising the base station, the relays and the
destination terminal. The complexity of the proposed RLS channel
estimation algorithm is $O(((n_r+1)ML))$.

\section{Simulations}

We evaluate the bit error rate (BER) performance of the proposed
joint power allocation and interference suppression (JPAIS)
algorithms and compare them with interference suppression schemes
without cooperation (NCIS) \cite{verdu} and with cooperation (CIS)
using an equal power allocation across the relays
\cite{venturino}. We consider a stationary DS-CDMA network with
randomly generated spreading codes with a processing gain $N=16$.
The block fading channels are generated considering a random power
delay profile with gains taken from a complex Gaussian variable
with unit variance and mean zero, $L=3$ paths, and are normalized
so that over the packets we have $E[{\boldsymbol {\mathcal
H}}^H[i]{\boldsymbol {\mathcal H}}[i]]=1$. We adopt the AF
cooperative strategy with repetitions and all the relays and the
destination terminal are equipped with linear MMSE receivers.It
should be remarked that the noise amplification of the AF protocol
is considered \cite{laneman04}. The receivers have either full
knowledge of the channel and the noise variance or are adaptive
and estimate all the required coefficients and the channels using
the proposed SG and RLS algorithms with optimized parameters. We
employ packets with $1500$ QPSK symbols and average the curves
over $1000$ runs. For the adaptive receivers, we provide training
sequences with $200$ symbols placed at the preamble of the
packets. After the training sequence, the adaptive receivers are
switched to decision-directed mode.

\begin{figure}[!htb]
\begin{center}
\def\epsfsize#1#2{1\columnwidth}
\epsfbox{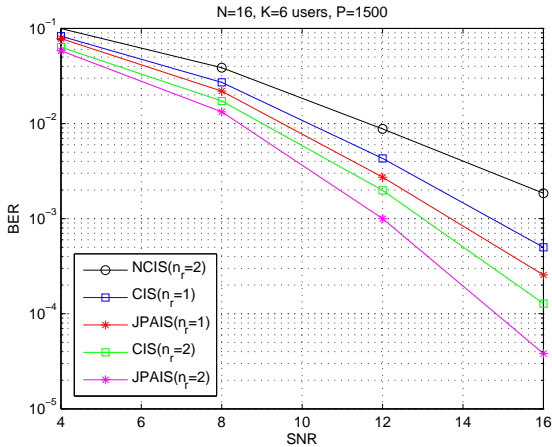} \caption{BER performance versus SNR for the
optimal linear MMSE detectors. Parameters: $\lambda =0.02$.}
\label{fig1}
\end{center}
\end{figure}

\begin{figure}[!htb]
\begin{center}
\def\epsfsize#1#2{1\columnwidth}
\epsfbox{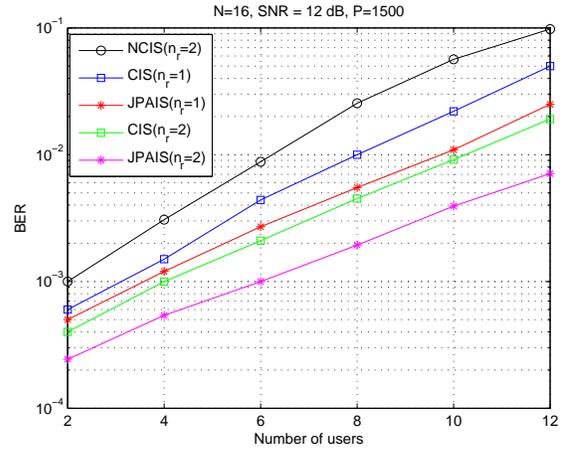} \caption{BER performance versus number of users
for the optimal linear MMSE detectors. Parameters: $\lambda =0.02$.}
\label{fig2}
\end{center}
\end{figure}

In the first experiment, we consider the proposed joint power
allocation and interference suppression (JPAIS) method with the
MMSE expressions of (\ref{wvec}) and (\ref{avec}). We compare the
proposed scheme with a non-cooperative approach (NCIS) and a
cooperative scheme with equal power allocation (CIS) for $n_r=1,2$
relays. The results shown in Figs. \ref{fig1} and \ref{fig2}
illustrate the performance improvement achieved by the proposed
JPAIS scheme, which significantly outperforms the CIS and the NCIS
techniques. As the number of relays is increased so is the
performance, reflecting the exploitation of the spatial diversity.
In the scenario studied, the proposed JPAIS approach can
accommodate up to $3$ more users as compared to the CIS scheme and
double the capacity as compared with the NCIS for the same BER
performance.

\begin{figure}[!htb]
\begin{center}
\def\epsfsize#1#2{1\columnwidth}
\epsfbox{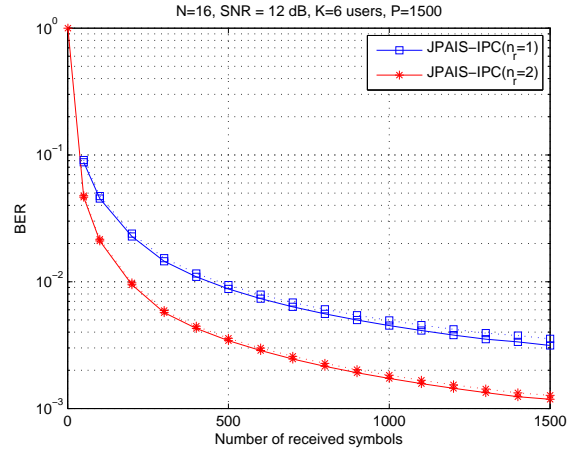} \caption{BER performance versus number of
symbols. The curves for the adaptive SG algorithms using the
solutions for the Lagrange multipliers are in solid lines, whereas
those of the adaptive SG algorithms with a simple normalization are
in dotted lines. Parameters: $\lambda=0.02$ (for MMSE schemes),
$\mu=0.025$, $\alpha=0.015$, $\nu = 0.01$ (for adaptive schemes).}
\label{fig3x}
\end{center}
\end{figure}

The second experiment depicted in Fig. \ref{fig3x} shows the BER
performance of the proposed adaptive SG algorithms (JPAIS) against
the existing NCIS and CIS schemes with $n_r=1$ and $n_r=2$ relays.
The aim of this experiment is to compare the proposed SG
algorithms that utilize the solutions for the Lagrange multipliers
in the recursions (\label{sgrew} \label{sgreca}) with the simpler
method that introduces a normalization of the power allocation
vector (\label{sgrew}, \label{sgreca2} and \label{sgreca2x}). The
techniques compared employ SG algorithms for estimation of the
coefficients of the channel. From the results we notice that the
SG recursions that obtain the values of the Lagrange multipliers
via the solution of the quadratic equation have a slightly better
performance than the normalization-based approach. This is
basically due to a greater precision in the computation of the
power allocation coefficients. For this reason, we will adopt this
version for the remaining experiments.

\begin{figure}[!htb]
\begin{center}
\def\epsfsize#1#2{1\columnwidth}
\epsfbox{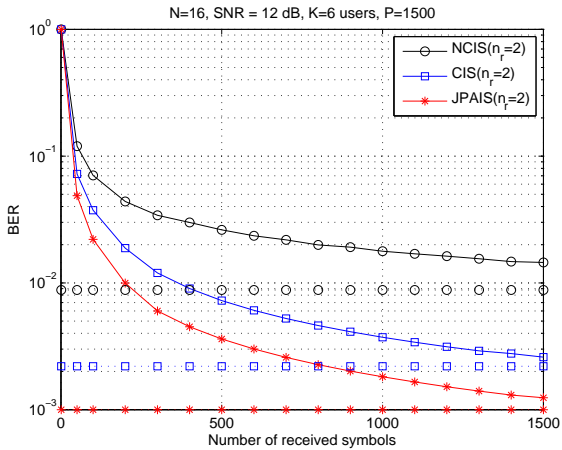} \caption{BER performance versus number of
symbols. The curves for the adaptive SG algorithms are in solid
lines, whereas those of the optimal MMSE schemes are in dotted
lines. Parameters: $\lambda=0.02$ (for MMSE schemes), $\mu=0.025$,
$\alpha=0.015$, $\nu = 0.01$ (for adaptive schemes).} \label{fig3}
\end{center}
\end{figure}

The third experiment depicted in Fig. \ref{fig3} shows the BER
performance of the proposed adaptive SG algorithms (JPAIS) against
the existing NCIS and CIS schemes with $n_r=2$ relays. All
techniques employ SG algorithms for estimation of the coefficients
of the channel, the receiver filters and the power allocation for
each user (JPAIS only). The complexity of the proposed algorithms
is linear with the filter length of the receivers times the number
of relays $n_r$, whereas the optimal MMSE schemes require cubic
complexity. From the results, we can verify that the proposed
adaptive estimation algorithms converge to approximately the same
level of the MMSE schemes, which have full channel and noise
variance knowledge.

\begin{figure}[!htb]
\begin{center}
\def\epsfsize#1#2{1\columnwidth}
\epsfbox{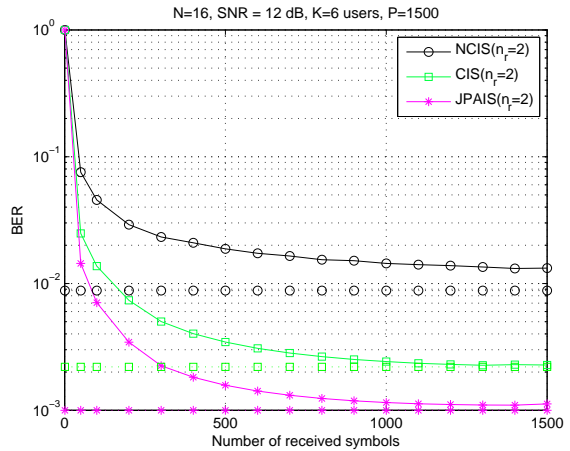} \caption{BER performance versus number of
symbols. The curves for the adaptive schemes are in solid lines,
whereas those of the optimal MMSE schemes are in dotted lines.
Parameters: $\alpha=0.998$.} \label{fig4}
\end{center}
\end{figure}

The fourth experiment depicted in Fig. \ref{fig4} shows the BER
performance of the proposed adaptive algorithms (JPAIS) against
the existing NCIS and CIS schemes with $n_r=2$ relays. The
techniques NCIS and CIS employ RLS algorithms for estimation of
the coefficients of the channel and the receiver. The proposed
JPAIS scheme and RLS algorithms estimates the parameters of the
channel, the receiver and the power allocation. The complexity of
the proposed adaptive algorithms is quadratic with the filter
length of the receivers and the number of relays $n_r$, whereas
the proposed optimal MMSE schemes require cubic complexity. From
the results, we can verify that the proposed adaptive RLS
estimation algorithms converge to approximately the same level of
the MMSE schemes, which have full channel and noise variance
knowledge. This indicates that the proposed RLS algorithms work
very well and can determine the coefficients of the channels and
the receivers.

\section{Conclusions}

This paper presented joint power allocation and interference
mitigation techniques for the downlink of spread spectrum systems
which employ multiple relays and the AF cooperation strategy. A
joint constrained optimization framework that considers the
allocation of power levels across the relays subject to an
individual power constraint and the design of linear receivers for
interference suppression was presented. We then derived MMSE
expressions and SG and RLS algorithms for determining the power
allocation and the parameters of the receiver. We also developed
SG and RLS channel estimation algorithms were also developed to
compute the coefficients of the channels across the base station,
the relays and the destination terminal. The simulations showed
that the proposed algorithms can obtain significant gains in
performance and capacity over non-cooperative systems and
cooperative schemes with equal power allocation. Future work will
consider a study of the proposed algorithms and their extension to
MIMO, OFDM systems, time-varying channels and limited feedback
issues.


\begin{thebibliography}{100}



{

\bibitem{sendonaris}
A. Sendonaris, E. Erkip, and B. Aazhang, "User cooperation
diversity - Parts I and II," \textit{IEEE Trans. Commun.}, vol.
51, no. 11, pp. 1927-1948, November 2003.

\bibitem{laneman03}
J. N. Laneman and G. W. Wornell, "Distributed space-time-coded
protocols for exploiting cooperative diversity in wireless
networks," \textit{IEEE Trans. Inf. Theory}, vol. 49, no. 10, pp.
2415-2425, Oct. 2003.

\bibitem{laneman04} J. N. Laneman and G. W. Wornell, "Cooperative diversity in
wireless networks: Efficient protocols and outage behaviour,"
\textit{IEEE Trans. Inf. Theory}, vol. 50, no. 12, pp. 3062-3080,
Dec. 2004.

\bibitem{huang}
W. J. Huang, Y. W. Hong and C. C. J. Kuo, ``Decode-and-forward
cooperative relay with multi-user detection in uplink CDMA
networks," in \textit{Proc. IEEE Global Telecommunications
Conference}, November 2007, pp. 4397-4401.

\bibitem{kramer}
G. Kramer, M. Gastpar and P. Gupta, ``Cooperative strategies and
capacity theorems for relay networks," \textit{IEEE Trans. Inf.
Theory}, vol. 51, no. 9, pp. 3037-3063, September 2005.

\bibitem{luo}
J. Luo, R. S. Blum, L. J.  Cimini, L. J   Greenstein, A. M.
Haimovich, ``Decode-and-Forward Cooperative Diversity with Power
Allocation in Wireless Networks", \textit{IEEE Transactions on
Wireless Communications}, vol. 6, no. 3, pp. 793 - 799, March
2007.

\bibitem{long}
L. Long and E. Hossain, ``Cross-layer optimization frameworks for
multihop wireless networks using cooperative diversity``,
\textit{IEEE Transactions on Wireless Communications}, vol. 7, no.
7, pp. 2592-2602, July 2008.

\bibitem{venturino} L. Venturino, X. Wang and M.
Lops, ``Multiuser detection for cooperative networks and
performance analysis," \textit{IEEE Trans. Sig. Proc.}, vol. 54,
no. 9, September 2006.

\bibitem{vardhe}
K. Vardhe, D. Reynolds, M. C. Valenti, ``The performance of
multi-user cooperative diversity in an asynchronous CDMA uplink",
\textit{IEEE Transactions on Wireless Communications}, vol. 7, no.
5, Part 2, May 2008, pp. 1930 - 1940.

\bibitem{souryal}
M. R. Souryal, ; B. R. Vojcic, ; L. Pickholtz, ``Adaptive
modulation in ad hoc DS/CDMA packet radio networks", \textit{IEEE
Transactions on Communications},  vol. 54,  no. 4,  April 2006 pp.
714 - 725.

\bibitem{comaniciu}
C. Comaniciu, and H. V. Poor, ``On the capacity of mobile ad hoc
networks with delay constraints" \textit{IEEE Transactions on
Wireless Communications}, vol. 5, no. 8, August 2006, pp.2061 -
2071.

\bibitem{tds1}
P. Clarke and R. C. de Lamare, ``Joint Transmit Diversity
Optimization and Relay Selection for Multi-relay Cooperative MIMO
Systems Using Discrete Stochastic Algorithms", IEEE Communications
Letters, vol. 15, no. 10, 2011,  pp.1035-1037.

\bibitem{tds2}
P. Clarke ad R. C. de Lamare, "Transmit Diversity and Relay
Selection Algorithms for Multi-relay Cooperative MIMO Systems", IEEE
Transactions on Vehicular Technology, vol. 61 , no. 3, March 2012,
Page(s): 1084 - 1098.

\bibitem{levorato}
M. Levorato, S. Tomasin, M. Zorzi, ``Cooperative spatial
multiplexing for ad hoc networks with hybrid ARQ: system design
and performance analysis", \textit{IEEE Transactions on
Communications}, vol. 56, no. 9, September 2008. pp. 1545 - 1555.

\bibitem{fischione}
C.Fischione, K. H. Johansson,  A. Sangiovanni-Vincentelli, B.
Zurita Ares, ``Minimum Energy coding in CDMA Wireless Sensor
Networks," \textit{IEEE Transactions on Wireless Communications},
vol. 8, no. 2, Feb. 2009, pp. 985 - 994.

\bibitem{kastrinogiannis}
T. Kastrinogiannis, V. Karyotis, S. Papavassiliou, ``An
Opportunistic Combined Power and Rate Allocation Approach in CDMA
Ad Hoc Networks",  2008 IEEE Sarnoff Symposium, 28-30 April 2008,
pp. 1 - 5.


\bibitem{chliu}
Chun-Hung Liu, ``Energy-Optimized Low-Complexity Control of Power
and Rate in Clustered CDMA Sensor Networks with Multirate
Constraints", Proc. IEEE 66th Vehicular Technology Conference, 30
Sept. - 3 Oct. 2007, pp. 331 - 335.

\bibitem{chen}
Min Chen; Changyoon Oh; Yener, A ,``Efficient Scheduling for Delay
Constrained CDMA Wireless Sensor Networks", Proc. IEEE 64th
Vehicular Technology Conference, VTC-2006 Fall. 2006, 25-28
September 2006, pp. 1 - 5.

\bibitem{delamare_iswcs}
R. C. de Lamare, ``Joint Power Allocation and Interference
Mitigation Techniques for Cooperative Spread Spectrum Systems with
Multiple Relays", Proc. IEEE International Symposium on Wireless
Communications Systems, 2009.

\bibitem{jpais_iet}
R. C. de Lamare, ``Joint iterative power allocation and linear
interference suppression algorithms for cooperative DS-CDMA
networks", IET Communications, vol. 6, no. 13 , 2012, pp. 1930-1942.

\bibitem{verdu}
S. Verdu, {\it Multiuser Detection}, Cambridge, 1998.

\bibitem{delamaresp}
R. C. de Lamare and R. Sampaio-Neto, ``Adaptive Reduced-Rank MMSE
Filtering with Interpolated FIR Filters and Adaptive Interpolators",
\textit{IEEE Signal Processing Letters}, vol. 12, no. 3, March,
2005.

\bibitem{delamarecl}
R. C. de Lamare and Raimundo Sampaio-Neto, ``Reduced-rank
Interference Suppression for DS-CDMA based on Interpolated FIR
Filters", \textit{IEEE Communications Letters}, vol. 9, no. 3, March
2005.

\bibitem{delamaretvt}
R. C. de Lamare and R. Sampaio-Neto, ``Adaptive Interference
Suppression for DS-CDMA Systems based on Interpolated FIR Filters
with Adaptive Interpolators in Multipath Channels", \textit{IEEE
Trans. Vehicular Technology}, Vol. 56, no. 6, September 2007, 2457
- 2474.

\bibitem{delamarespl07}
R. C. de Lamare and R. Sampaio-Neto, ``Reduced-Rank Adaptive
Filtering Based on Joint Iterative Optimization of Adaptive
Filters", \textit{IEEE Signal Processing Letters}, Vol. 14, no. 12,
December 2007.

\bibitem{delamaretvt10}
R. C. de Lamare and R. Sampaio-Neto, ``Reduced-Rank Space-Time
Adaptive Interference Suppression With Joint Iterative Least Squares
Algorithms for Spread-Spectrum Systems," \textit{IEEE Transactions
on Vehicular Technology}, vol.59, no.3, March 2010, pp.1217-1228.

\bibitem{jidf_icassp}
R. C. de Lamare and R. Sampaio-Neto, ``Adaptive Reduced-Rank MMSE
Parameter Estimation based on an Adaptive Diversity Combined
Decimation and Interpolation Scheme," \textit{Proc. IEEE
International Conference on Acoustics, Speech and Signal
Processing}, April 15-20, 2007, vol. 3, pp. III-1317-III-1320.

\bibitem{jidf}
R. C. de Lamare and R. Sampaio-Neto, ``Adaptive Reduced-Rank
Processing Based on Joint and Iterative Interpolation, Decimation,
and Filtering," \textit{IEEE Transactions on Signal Processing},
vol. 57,  no. 7,  July 2009, pp. 2503 - 2514.

\bibitem{barc}
R.C. de Lamare, R. Sampaio-Neto and M. Haardt, "Blind Adaptive
Constrained Constant-Modulus Reduced-Rank Interference Suppression
Algorithms Based on Interpolation and Switched Decimation,"
\textit{IEEE Trans. on Signal Processing},  vol.59, no.2,
pp.681-695, Feb. 2011.

\bibitem{delamaretc}
R. C. de Lamare and R. Sampaio-Neto, Minimum Mean Squared Error
Iterative Successive Parallel Arbitrated Decision Feedback
Detectors for DS-CDMA Systems," \textit{IEEE Transactions on
Communications}, vol. 56, no. 5, May 2008, pp. 778 - 789.

\bibitem{delamare_itic}
R.C. de Lamare, R. Sampaio-Neto, A. Hjorungnes, ``Joint iterative
interference cancellation and parameter estimation for cdma
systems", \textit{IEEE Communications Letters}, vol. 11, no. 12,
December 2007, pp. 916 - 918.

\bibitem{haykin}
S. Haykin, \textit{ Adaptive Filter Theory}, 4th ed. Englewood
Cliffs, NJ: Prentice- Hall, 2002.

\bibitem{golub}
G. H. Golub and C. F. van Loan, \textit{ Matrix Computations}, 3rd
ed., The Johns Hopkins University Press, Baltimore, Md, 1996.


\bibitem{rappa}
T. S. Rappaport, {\it Wireless Communications}, Prentice-Hall,
Englewood Cliffs, NJ, 1996.}


\end{thebibliography}
\end{document}